\documentclass[%
 reprint,onecolumn,
nofootinbib,
 amsmath,amssymb,
 aps,
]{revtex4-2}

\usepackage{graphicx}
\usepackage{dcolumn}
\usepackage{bm}
\usepackage{xcolor}
\usepackage{hyperref}
\usepackage{cancel}

\begin{document}

\newtheorem{theorem}{Theorem}
\newtheorem{definition}{Definition}
\newtheorem{lemma}{Lemma}
\newtheorem{proposition}{Proposition}
\newtheorem{remark}{Remark}
\newtheorem{corollary}{Corollary}
\newtheorem{example}{Example}


\title{Generalized diffusion process with nonlocal interactions: Continuous time random walk model and stochastic resetting
}


\author{Pece Trajanovski\textsuperscript{1,2}, Irina Petreska\textsuperscript{2}, Katarzyna G\'orska\textsuperscript{3}, Ljupco Kocarev\textsuperscript{1,4}, Trifce Sandev\textsuperscript{1,2,5}}
\address{\textsuperscript{1}\textit{Research Center for Computer Science and Information Technologies, Macedonian Academy of Sciences and Arts, Bul. Krste Misirkov 2, 1000 Skopje, Macedonia}}
\address{\textsuperscript{2}\textit{Institute of Physics, Faculty of Natural Sciences and Mathematics, Ss.~Cyril and Methodius University, Arhimedova 3, 1000 Skopje, Macedonia}}
\address{\textsuperscript{3}Institute of Nuclear Physics, Polish Academy of Sciences, ul.~Radzikowskiego 152, PL-31342 Krak\'ow, Poland}
\address{\textsuperscript{4}\textit{Faculty of Computer Science and Engineering, Ss. Cyril and Methodius University, PO Box 393, 1000 Skopje, Macedonia}}
\address{\textsuperscript{5}\textit{Department of Physics, Korea University, Seoul 02841, Korea}}

\date{\today}
\begin{abstract}
A space fractional diffusion-like equation is introduced, which embodies the nonlocality in time, represented by the memory kernel and the non-locality in space. A specific example of the nonlocal term is considered in combination with three different forms of the memory kernel. To analyse the probability density function, we utilize the subordination approach. Subsequently, the corresponding continuous time random walk model is presented. Furthermore, we investigate the effects of the stochastic resetting on the dynamics of the process and we showed that in the long time limit the system approaches a nonequilibrium stationary state.
\end{abstract}

\maketitle

\section{Introduction}

One of the most intriguing questions in modeling of the diffusion processes is how to incorporate the inter-particle interactions, especially when the nonlocal effects are important. Depending on the size and complexity of the system, the local, i.e., averaged macroscale description of the diffusion processes is not always sufficient. In complex systems, that manifest heterogeneity caused by porosity or anisotropy, the nonlocal effects play a crucial role in the diffusion processes, as the time intervals and the distances of interest are comparable to the characteristic diffusion parameters, such as the mean free path and the timescale of the particles' velocity variations~\cite{nonlocalkoch}. In general, nonlocality is related to long-range interactions, which means that the motion of one particle is not only affected by its neighbours, but also by other particles that are far from the observed particle~\cite{nonlocalbalague}. Therefore, taking into account these interactions is crucial in explaining a plethora of collective phenomena, ranging from nonlocal correlations in quantum systems to population dynamics in ecosystem and the evolution of species itself~\cite{nonlocalintro1, nature0, nature, PRLnonlocal, repprog, plos1, plos2}. 

In order to include the nonlocal interactions, the commonly used local potentials should be substituted by a nonlocal form, that involves the dependence on the interparticle distance, represented as $V(x-\xi)$ and enters as a kernel in the convolution integral with the probability denisty function. Such nonlocal potentials in the quantum context were previously considered in \cite{lenzi epjb,jiang,jmp2014,lenzi book}. By analysing the Green's functions, it was shown that the anomalous behaviour of the probability density function can be adequately described by the introduction of a nonlocal potential in the Schr\"{o}dinger equation. The classical diffusion equation in presence of a nonlocal term was considered in \cite{lenzi AST}.

Strictly speaking, the nonlocality is not only associated with the spatial correlations, but it is also related to the temporal relations between the events, which lead to non-Markovianity in many of the real life diffusion processes. Inclusion of the nonlocality in time in the mathematical diffusion models can be achieved by using the time dependent memory kernel~\cite{sokolov,sokolov_book}.

In this work, we propose a generalised diffusion-like equation which embodies the nonlocality in time, represented by the memory effects and the non-locality in space in a single model. Further generalisation is achieved by using the space-fractional Riesz derivative instead of the Laplacian. Along with the analysis of the different forms of the memory kernel and the nonlocal potential, we extend our investigation further by considering also the effects of the stochastic resetting by means of the renewal equation approach.

The paper is organized as follows. In Section~\ref{Sec2}, the foundations of the model and its solution are represented. In Section~\ref{Sec3}, the probability density function is analysed from the contionious time random walk. In Section~\ref{Sec4}, we introduce a stochastic ressetting in our model, while in Section~\ref{Sec5}, one special case of the nonlocal interaction in presence of three different forms of the memory kernel is considered. At the end, in Section~\ref{sec_summary} we give the summary, as well as the directions for future research.

\section{The Model and Its Solution}\label{Sec2}

In this paper we introduce the following diffusion-like equation in presence of a nonlocal term
\begin{equation}\label{memory diffusion eq1}
\int_{0}^{t}\gamma(t-\tau)\frac{\partial}{\partial\tau}u(x,\tau)\,d\tau=\mathcal{D}_{\alpha,\gamma}\frac{\partial^{\alpha}}{\partial |x|^{\alpha}}u(x,t)-\int_{-\infty}^{\infty}V(x-\xi)u(\xi,t)\,d\xi,
\end{equation}
where $u(x,t)$ is the probability density function (PDF), $\mathcal{D}_{\alpha,\gamma}$ is the generalised diffusion coefficient, $\gamma(t)$ is the memory kernel, $\frac{\partial^{\alpha}}{\partial|x|^{\alpha}}$ is the Riesz fractional derivative of order $1<\alpha\leq2$, and the second term form the right hand side of the equation gives the nonlocal effect in space represented by the kernel function $V(x)$. The Riesz derivative is given as a pseudo-differential operator with the Fourier symbol
$-|\kappa|^\alpha$, $\kappa\in \mathbb{R}$~\cite{feller}
\begin{equation}\label{riesz}
\frac{\partial^\alpha}{\partial|x|^\alpha}f(x)=\mathcal{F}^{-1}\left[-|\kappa|^\alpha
\tilde{F}(\kappa)\right](x),
\end{equation}
where $\tilde{F}(\kappa)=\mathcal{F}\left[f(x)\right](\kappa)=\int_{-\infty}^{\infty}f(x)e^{\imath\kappa x}dx$ is the Fourier transform, $\mathcal{F}^{-1}\left[\tilde{F}(\kappa)\right](x)=\frac{1}{2\pi}\int_{-\infty}^{\infty}f(x)e^{-\imath\kappa x}d\kappa$ is the inverse Fourier transform, $\mathcal{D}_{\alpha}$ is a generalized diffusion coefficient with physical dimension $\left[D_{\alpha}\right]$ which depends on the form of memory kernel $\gamma(t)$, and the last term is the nonlocal term. Such nonlocal term has been used in different forms of the Schr\"{o}dinger equation \cite{lenzi epjb,jiang,jmp2014} and diffusion equations \cite{lenzi AST} obtaining various anomalous diffusive behaviour.

In absence of the nonlocal term ($V(x)=0$), eq.~(\ref{memory diffusion eq1}) becomes the following generalised diffusion equation with memory~\cite{jstat}
\begin{equation}\label{memory diffusion eq V0}
\int_{0}^{t}\gamma(t-\tau)\frac{\partial}{\partial\tau}u(x,\tau)\,d\tau=\mathcal{D}_{\alpha,\gamma}\frac{\partial^{\alpha}}{\partial |x|^{\alpha}}u(x,t).
\end{equation}
This equation in case of Dirac delta memory kernel $\gamma(t)=\delta(t)$ corresponds to the space fractional diffusion equation for L\'{e}vy flights with L\'{e}vy index $\alpha$, i.e.,
\begin{equation}
    \frac{\partial}{\partial t}u(x,t)=\mathcal{D}_{\alpha}\frac{\partial^{\alpha}}{\partial |x|^{\alpha}}u(x,t),
\end{equation}
where the diffusion coefficient is of dimension $[\mathcal{D}_{\alpha}]=\mathrm{m}^{\alpha}\mathrm{s}^{-1}$. In case of power-law memory kernel $\gamma(t)=\frac{t^{-\mu}}{\Gamma(1-\mu)}$ ($0<\mu\leq1$) we obtain the space-time fractional diffusion equation 
\begin{equation}
{_{t}}D_{0+}^{\mu}u(x,t)=\mathcal{D}_{\alpha,\mu}\,\frac{\partial^{\alpha}}{\partial |x|^{\alpha}}u(x,t),
\end{equation}
where the diffusion coefficient is of dimension $[\mathcal{D}_{\alpha,\mu}]=\mathrm{m}^{\alpha}\mathrm{s}^{-\mu}$ and 
${_{t}}D_{0+}^{\mu}f(t)$ is the Caputo fractional derivative 
of order $0<\mu<1$~\cite{prudnikov} 
\begin{equation}\label{caputo derivative}
{_{\mathrm{C}}}D_{t}^{\mu}f(t)=\frac{1}{\Gamma(1-\alpha)}\int_{0}^{t}(t-t')^{-\mu}\frac{d}{dt'}f(t')\,dt'.
\end{equation}
This equation gives same results for the PDF as those obtained from the subdiffusive continuous time random walks for L\'{e}vy flights. Such fractional diffusion equations have been of current interest for modeling anomalous diffusive processes \cite{metzler report,pagnini3,pagnini,pagnini1,pagnini2,mainardi_et_al2}.

Here we note that the diffusion-like equation (\ref{memory diffusion eq V0}) can be transformed to the natural form of distributed order diffusion equation since if we consider distributed order memory kernel of form \cite{kochubei}
\begin{equation}\label{memory kernel_distributed}
\gamma(t)=\int_{0}^{1}\tau^{\bar{\gamma}-1}p(\bar{\gamma})\frac{t^{-\bar{\gamma}}}{\Gamma(1-\bar{\gamma})}d\bar{\gamma},
\end{equation}
where $p(\bar{\gamma})$ is a weight function with $\int_{0}^{1}p(\bar{\gamma})d\bar{\gamma}=1$. We note that for $p(\bar{\gamma})=1$, the memory kernel is so-called uniformly distributed memory kernel. For such distributed order memory kernels eq.~(\ref{memory diffusion eq V0}) becomes

\begin{equation}\label{distributed diffusion eq}
\int_{0}^{t}\left[\int_{0}^{1}\tau^{\bar{\gamma}-1}p(\bar{\gamma})\frac{(t-\tau)^{-\bar{\gamma}}}{\Gamma(1-\bar{\gamma})}
\frac{\partial}{\partial\tau}u(x,\tau)\,d\bar{\gamma}\right]d\tau=\mathcal{D}_{\alpha,\gamma}\frac{\partial^{\alpha}}{\partial |x|^{\alpha}}u(x,t),
\end{equation}
i.e.,~\cite{chechkin1,chechkin11,chechkin12}
\begin{equation}\label{distributed diffusion eq_2}
\int_{0}^{1}\tau^{\bar{\gamma}-1}p(\bar{\gamma}){_{t}}D_{0+}^{\bar{\gamma}}u(x,\tau)\,d\bar{\gamma}=\mathcal{D}_{\alpha,\gamma}\frac{\partial^{\alpha}}{\partial |x|^{\alpha}}u(x,t),
\end{equation}
where the diffusion coefficient is of dimension $[\mathcal{D}_{\alpha,\gamma}]=\mathrm{m}^{\alpha}\mathrm{s}^{-1}$

Diffusion-like equation (\ref{memory diffusion eq V0}) with $\alpha=2$ was considered in~\cite{tateishi}, while Sokolov and Klafter \cite{sokolov} obtained such integral equation for $\alpha=2$ for PDF in the analysis of anomalous diffusive process subordinated to normal diffusion under operational time, where the memory kernel is connected to the cumulative distribution function of waiting times, i.e., it is a probability to make no step up to time $t$. The Schr\"{o}dinger equation with nonlocal term has been analysed in~\cite{jmp2014} where the Green function approach has been employed for analysis of the particle dynamics.

First, we apply Laplace transform (the Laplace transform is defined by $\hat{F}(s)=\mathcal{L}[f(t)]=\int_{0}^{\infty}f(t)e^{-st}dt$) of equation (\ref{memory diffusion eq}). Thus, one obtains
\begin{equation}\label{memory diffusion eq laplace}
\hat{\gamma}(s)\left[s\hat{U}(x,s)-u(x,0)\right]=\mathcal{D}_{\alpha,\gamma}\frac{\partial^{\alpha}}{\partial |x|^{\alpha}}\hat{U}(x,s)-\int_{-\infty}^{\infty}\hat{V}(x-\xi)\hat{U}(\xi,s)\,d\xi,
\end{equation}
where $\hat{U}(x,s)=\mathcal{L}\left[u(x,t)\right]$ and $u(x,0)$ is the initial condition. Next, by Fourier transform to eq.~(\ref{memory diffusion eq laplace}), by using the Fourier transform formula for convolution $\mathcal{F}\left[\int_{-\infty}^{\infty}f(x-\zeta)g(\zeta)\,d\zeta\right]=\mathcal{F}[f(x)]\mathcal{F}[g(x)]=\tilde{F}(\kappa)\tilde{G}(\kappa)$, it follows
\begin{equation}\label{memory diffusion eq laplace-fourier}
\hat{\gamma}(s)\left[s\tilde{\hat{U}}(\kappa,s)-\tilde{U}(\kappa,0)\right]
=-\mathcal{D}_{\alpha,\gamma}|\kappa|^{\alpha}\tilde{\hat{U}}(\kappa,s)-\tilde{V}(\kappa)\tilde{\hat{U}}(\kappa,s),
\end{equation}
where $\tilde{\hat{U}}(\kappa,s)=\mathcal{F}\left[\hat{U}(x,s)\right]$, $\tilde{V}(k)=\mathcal{F}\left[V(x)\right]$. Therefore, for the field variable in the Fourier-Laplace space, we obtain
\begin{equation}\label{memory diffusion eq laplace-fourier U}
\tilde{\hat{U}}(\kappa,s)=\frac{\hat{\gamma}(s)}{s\hat{\gamma}(s)+\mathcal{D}_{\alpha,\gamma}|\kappa|^{\alpha}+\tilde{V}(\kappa)}\tilde{U}(\kappa,0).
\end{equation}
In absence of the nonlocal term, one finds
\begin{equation}\label{memory diffusion eq laplace-fourier U V0}
\tilde{\hat{U}}(\kappa,s)=\frac{\hat{\gamma}(s)}{s\hat{\gamma}(s)+\mathcal{D}_{\alpha,\gamma}|\kappa|^{\alpha}}\tilde{U}(\kappa,0).
\end{equation}

Let us discuss the case where the nonlocal term $V(x)=K(x)$ fulfills $\tilde{K}(\kappa=0)=0$ for delta initial condition $u(x,0)=\delta(x)$. Therefore, Eq.~(\ref{memory diffusion eq laplace-fourier U}) becomes
\begin{equation}\label{memory diffusion eq laplace-fourier U2}
\tilde{\hat{U}}(\kappa,s)=\frac{\hat{\gamma}(s)}{s\hat{\gamma}(s)+\mathcal{D}_{\alpha,\gamma}|\kappa|^{\alpha}+\tilde{K}(\kappa)}.
\end{equation}
From here one concludes that the solution $u(x,t)$ is normalized to 1 since $\tilde{K}(\kappa=0)=0$, and thus $\tilde{\hat{U}}(\kappa=0,s)=1/s$, so in this case the solution may behave as PDF.

\begin{remark}[{\bf Solution for Dirac delta memory kernel}]\label{remark1}

For $\gamma(t)=\delta(t)$ for the solution~(\ref{memory diffusion eq laplace-fourier U2}), we have
\begin{equation}\label{memory diffusion eq laplace-fourier U2ex1}
u(x,t)=\mathcal{F}^{-1}\left[\mathcal{L}^{-1}\left[\frac{1}{s+\mathcal{D}_{\alpha,\gamma}|\kappa|^{\alpha}+\tilde{K}(\kappa)}\right]\right]=\mathcal{F}^{-1}\left[e^{-\left(\mathcal{D}_{\alpha,\gamma}|\kappa|^{\alpha}+\tilde{K}(\kappa)\right)t}\right].
\end{equation}
    
\end{remark}

\begin{remark}[{\bf Solution for power-law memory kernel}]

For $\gamma(t)=\frac{t^{-\mu}}{\Gamma(1-\mu)}$, $0<\mu<1$ for the solution~(\ref{memory diffusion eq laplace-fourier U2}), we have
\begin{equation}\label{memory diffusion eq laplace-fourier U2ex2}
u(x,t)=\mathcal{F}^{-1}\left[\mathcal{L}^{-1}\left[\frac{s^{\mu-1}}{s^\mu+\mathcal{D}_{\alpha,\gamma}|\kappa|^{\alpha}+\tilde{K}(\kappa)}\right]\right]=\mathcal{F}^{-1}\left[E_{\mu}\left(-\left[\mathcal{D}_{\alpha,\gamma}|\kappa|^{\alpha}+\tilde{K}(\kappa)\right]t^{\mu}\right)\right],
\end{equation}
where 
\begin{equation}\label{one parameter ML}
E_{\alpha}(z)=\sum_{k=0}^{\infty}\frac{z^k}{\Gamma(\alpha k+\beta)}, \quad z\in C, \quad \Re\{\alpha\}>0,
\end{equation}
is the one parameter Mittag-Leffler function~\cite{ml}. In eq.~(\ref{memory diffusion eq laplace-fourier U2ex2}) we use the Laplace transform formula for the one parameter Mittag-Leffler function~\cite{ml}
\begin{equation}\label{Laplace ML1}
\mathcal{L}\left[E_{\alpha}(\mp\lambda t^{\alpha})\right]=\frac{s^{\alpha-1}}{s^\alpha\pm\lambda},
\end{equation}
where $|\lambda/s^{\alpha}|<1$.
\end{remark}

\begin{remark}[{\bf Subordination~\cite{sokolov,metzler report,bazhlekova,book_ws,kg_fcaa}}]\label{remark3} Here we note that the PDF of the generalised equation~(\ref{memory diffusion eq1}) can be found by using the subordination approach. Let us consider the following equation
\begin{equation}\label{memory diffusion eq1_standard}
\frac{\partial}{\partial t}u_0(x,t)=\mathcal{D}_{\alpha,\gamma}\frac{\partial^{\alpha}}{\partial |x|^{\alpha}}u_0(x,t)-\int_{-\infty}^{\infty}V(x-\xi)u_0(\xi,t)\,d\xi.
\end{equation}
Its Laplace transform reads
\begin{equation}\label{memory diffusion eq laplace standard}
s\hat{U}_0(x,s)-u_0(x,0)=\mathcal{D}_{\alpha,\gamma}\frac{\partial^{\alpha}}{\partial |x|^{\alpha}}\hat{U}_0(x,s)-\int_{-\infty}^{\infty}\hat{V}(x-\xi)\hat{U}_0(\xi,s)\,d\xi.
\end{equation}
We introduce the change of the variables $s \rightarrow s\hat{\gamma}(s)$ and obtain
\begin{equation}\label{memory diffusion eq laplace standard2}
s\hat{\gamma}(s)\hat{U}_0(x,s\hat{\gamma}(s))-u_0(x,0)=\mathcal{D}_{\alpha,\gamma}\frac{\partial^{\alpha}}{\partial |x|^{\alpha}}\hat{U}_0(x,s\hat{\gamma}(s))-\int_{-\infty}^{\infty}\hat{V}(x-\xi)\hat{U}_0(\xi,s\hat{\gamma}(s))\,d\xi,
\end{equation}
which can be rewritten as
\begin{eqnarray}\label{memory diffusion eq laplace standard3}
&&s\left[\hat{\gamma}(s)\hat{U}_0(x,s\hat{\gamma}(s))\right]-u_0(x,0)=\frac{1}{\hat{\gamma}(s)}\left[\mathcal{D}_{\alpha,\gamma}\frac{\partial^{\alpha}}{\partial |x|^{\alpha}}\hat{\gamma}(s)\hat{U}_0(x,s\hat{\gamma}(s))-\int_{-\infty}^{\infty}\hat{V}(x-\xi)\hat{\gamma}(s)\hat{U}_0(\xi,s\hat{\gamma}(s))\,d\xi\right].
\end{eqnarray}
Let us introduce a new function
\begin{equation}\label{subordination_PDF}
    \hat{U}_{\mathrm{s}}(x,s)=\hat{\gamma}(s)\hat{U}_0(x,s\hat{\gamma}(s)).
\end{equation}
Thus, eq.~(\ref{memory diffusion eq laplace standard3}) becomes
\begin{equation}\label{memory diffusion eq laplace4}
\hat{\gamma}(s)\left[s\hat{U}_{\mathrm{s}}(x,s)-u(x,0)\right]=\mathcal{D}_{\alpha,\gamma}\frac{\partial^{\alpha}}{\partial |x|^{\alpha}}\hat{U}_{\mathrm{s}}(x,s)-\int_{-\infty}^{\infty}\hat{V}(x-\xi)\hat{U}_{\mathrm{s}}(\xi,s)\,d\xi,
\end{equation}
which by inverse Laplace transform yields eq.~(\ref{memory diffusion eq1}). Therefore, the PDF~(\ref{subordination_PDF}) corresponds to the solution of eq.~(\ref{memory diffusion eq1}), i.e., $u_{\mathrm{s}}(x,t)\equiv u(x,t)$. From eq.~(\ref{subordination_PDF}), one has
\begin{equation}\label{subordination_PDF2}
    \hat{U}_{\mathrm{s}}(x,s)=\hat{\gamma}(s)\hat{U}_0(x,s\hat{\gamma}(s))=\hat{\gamma}(s)\int_{0}^{\infty}u_0(x,\omega)e^{-\omega s\hat{\gamma}(s)}d\omega=\int_{0}^{\infty}u_0(x,\omega)\hat{h}(\omega,s)d\omega,
\end{equation}
where
\begin{equation}
    \hat{h}(\omega,s)=\hat{\gamma}(s)e^{-\omega s\hat{\gamma}(s)}
\end{equation}
is the so-called subordination function. Therefore, the PDF $u_{\mathrm{s}}(x,t)$ can be found if one knows the solution solution $u_0(x,t)$ of equation~(\ref{memory diffusion eq1_standard}) by using the subordination integral~(\ref{subordination_PDF2}), which by inverse Laplace transform becomes
\begin{equation}\label{subordination_PDFfinal}
    u_{\mathrm{s}}(x,t)=\int_{0}^{\infty}u_0(x,\omega)h(\omega,t)d\omega,
\end{equation}
where
\begin{equation}
    h(\omega,t)=\mathcal{L}^{-1}\left[\hat{\gamma}(s)e^{-\omega s\hat{\gamma}(s)}\right].
\end{equation}

\end{remark}

\section{CTRW Model}\label{Sec3}

Let us now analyse the PDF $W(x,t)$ of a particle being in position
$x$ at time $t$ from the CTRW model. PDF in the Fourier-Laplace
space is given by~\cite{metzler report}
\begin{equation}\label{CTRW}
\tilde{\hat{W}}(\kappa,s)=\frac{1-\hat{w}(s)}{s}\frac{\tilde{W}(\kappa,0)}{1-\tilde{\hat{\psi}}(\kappa,s)},
\end{equation}
where $\tilde{\hat{\psi}}(\kappa,s)$ is the joint PDF of jumps and
waiting times in the Fourier-Laplace space,
\begin{equation}\label{jump length}
\lambda(x)=\int_{0}^{\infty}\psi(x,t)\,dt
\end{equation}
is the jump length PDF,
\begin{equation}\label{waiting time}
w(t)=\int_{0}^{\infty}\psi(x,t)\,dx
\end{equation}
is the waiting time PDF, and $\tilde{W}(\kappa,0)$ is the Fourier transform of the initial condition $W(x,0)$. If the jump length and
waiting time are independent random variables, then we assume
decoupled joint PDF $\psi(x,t)=\lambda(x)w(t)$.

For the Poisson type waiting time PDF
$w(t)=\tau^{-1}e^{-\frac{t}{\tau}}$, which means that the
characteristic waiting time $T=\int_{0}^{\infty}tw(t)\,dt$ is
equal to $\tau$, and Gaussian type jump length PDF
$\lambda(x)=\frac{1}{\sqrt{4\pi\sigma^{2}}}e^{-\frac{x^{2}}{4\sigma^{2}}}$,
for which jump length variance
$\Sigma^{2}=\int_{-\infty}^{\infty}x^{2}\lambda(x)\,dx$ equals $\sigma^{2}$, for initial condition $W(x,0+)=\delta(x)$, from
relation (\ref{CTRW}) one obtains PDF for classical Brownian motion
\begin{equation}\label{CTRW Brownian}
\tilde{\hat{W}}(\kappa,s)=\frac{1}{s+\mathcal{D}_{2,1}\kappa^{2}},
\end{equation}
from where one may directly obtain the classical diffusion equation
with diffusion coefficient $\mathcal{D}_{2,1}=\sigma^{2}\tau^{-1}$ with dimension $\mathrm{m}^{2}\mathrm{s}^{-1}$.
Furthermore, if instead of Poisson type waiting time PDF one
consider long-tailed waiting time PDF of form
$w(t)\simeq(\tau/t)^{1+\mu}$, where $0<\mu<1$, for which
characteristic waiting time $T$ diverges, it can be shown that for initial condition $W(x,0+)=\delta(x)$, PDF in the Fourier-Laplace
space is given by 
\begin{equation}\label{CTRW Anomalous}
\tilde{\hat{W}}(\kappa,s)=\frac{s^{\mu-1}}{s^{\mu}+\mathcal{D}_{2,\mu}\kappa^{2}},
\end{equation}
where $\mathcal{D}_{2,\mu}=\sigma^{2}\tau^{-\mu}$ is the
generalized diffusion coefficient with dimension $\mathrm{m}^{2}\mathrm{s}^{-1}$. By applying inverse
Fourier-Laplace transform one obtains time fractional diffusion equation.

For the Poisson type waiting time PDF
$w(t)$, and L\'{e}vy distribution of the jump length
$\tilde{\lambda}(\kappa)=e^{-\sigma^{\alpha}|\kappa|^{\alpha}}\simeq1-\sigma^{\alpha}|\kappa|^{\alpha}$,
one can find PDF for L\'{e}vy flights
\begin{equation}\label{CTRW Levy}
\tilde{\hat{W}}(\kappa,s)=\frac{1}{s+\mathcal{D}_{\alpha,1}|\kappa|^{\alpha}},
\end{equation}
where diffusion coefficient $\mathcal{D}_{\alpha,1}=\sigma^{\alpha}\tau^{-1}$ is of dimension $\mathrm{m}^{\alpha}\mathrm{s}^{-1}$.

If one considers a CTRW characterized through broad PDFs for both waiting time and jump length, we arrive to the equation
\begin{equation}\label{CTRW Anomalous Levy}
\tilde{\hat{W}}(\kappa,s)=\frac{s^{\mu-1}}{s^{\mu}+\mathcal{D}_{\alpha,\mu}|\kappa|^{\alpha}},
\end{equation}
where $\mathcal{D}_{\alpha,\mu}=\sigma^{\alpha}\tau^{-\mu}$ with dimension $[\mathrm{m}^{\alpha}\mathrm{s}^{-\mu}]$.

Let us now consider a general form of the waiting time PDF in Laplace space~\cite{fcaa2015}
\begin{equation}\label{w(s) gamma(s)}
\hat{w}(s)=\frac{1}{1+\tau_{\gamma}s\hat{\gamma}(s)},
\end{equation}
where $\tau_{\gamma}$ has dimension which depends on the form of the
memory kernel $\gamma(t)$. Memory kernel is the same as one used in the diffusion-like equation~(\ref{memory diffusion eq V0}). The waiting time PDF (\ref{w(s) gamma(s)}) and L\'{e}vy type jump
length PDF, for $W(x,0)=\delta(x)$, yield the following form for
$\tilde{\hat{W}}(\kappa,s)$,
\begin{equation}\label{generalised diffusion eq}
\tilde{\hat{W}}(\kappa,s)=\frac{\hat{\gamma}(s)}{s\hat{\gamma}(s)+\mathcal{D}_{\alpha,\gamma}|\kappa|^{\alpha}},
\end{equation}
where $\mathcal{D}_{\alpha,\gamma}=\sigma^{\alpha}\tau_{\gamma}^{-1}$ has a dimension depending on the memory kernel. We note that this equation has a same form as eq.~(\ref{memory diffusion eq laplace-fourier U V0}). Thus, the considered
diffusion-like equation (\ref{memory diffusion eq V0}) with memory corresponds to the limit of CTRW with
waiting time PDF given by (\ref{w(s) gamma(s)}).

From eq.~(\ref{w(s) gamma(s)}), for the uniformly distributed order memory kernel, see eq.~(\ref{memory kernel_distributed}) with $p(\bar{\gamma})=1$, for which 
\begin{equation}\label{unif_distr_memory_laplace}
    \hat{\gamma}(s)=\frac{s-1}{s\log{s}},
\end{equation}
one finds the waiting time PDF in the long time limit~\cite{fcaa2015}
\begin{equation}\label{w(s) unif distr}
\hat{w}(t)=\mathcal{L}^{-1}\left[\frac{1}{1+\tau^{\bar{\gamma}-1}\frac{s-1}{\log{s}}}\right]\sim -\frac{d}{dt}\frac{1}{\log{t}}=\frac{1}{t\log^2{t}}.
\end{equation}

Next, we try to find the corresponding CTRW model to the generalised diffusion equation with nonlocal term~(\ref{memory diffusion eq1}). Let us consider a waiting time PDF~(\ref{w(s) gamma(s)}) 
and the following jump length PDF
\begin{equation}\label{jumpPDF}
    \tilde{\lambda}(\kappa)=e^{-\sigma^{\alpha}|\kappa|^{\alpha}-\tilde{M}(\kappa)}\simeq1-\left(\sigma^{\alpha}|\kappa|^{\alpha}+\tilde{M}(\kappa)\right),
\end{equation}
with $\tilde{M}(\kappa)\rightarrow0$ for $\kappa\rightarrow0$. From relation (\ref{CTRW}) we find the PDF in the Fourier-Laplace space
\begin{equation}\label{CTRW nonlocal delta(t)}
\tilde{\hat{W}}(\kappa,s)=\frac{\hat{\gamma}(s)}{s\hat{\gamma}(s)+\mathcal{D}_{\alpha,\gamma}|\kappa|^{\alpha}+\tilde{K}(\kappa)}\tilde{W}(\kappa,0),
\end{equation}
where $\mathcal{D}_{\alpha,\gamma}=\sigma^{\alpha}\tau_{\gamma}^{-1}$ and $\tilde{K}(\kappa)=\tilde{M}(\kappa)\tau_{\gamma}^{-1}$, and which is equivalent to eq.~(\ref{memory diffusion eq laplace-fourier U2}). 

Therefore, from~(\ref{jumpPDF}), by using the Fourier convolution, we can express the jump length PDF as a convolution integral
\begin{equation}\label{lambda_composed}
    \lambda(x) = \mathcal{F}^{-1}\left[e^{-\sigma^{\alpha}|\kappa|^{\alpha}-\tilde{K}(\kappa)}\right] = \int_{-\infty}^{\infty}\lambda_{1}(\alpha; \sigma, x-\xi)\lambda_{2}(\xi)\,d\xi,
\end{equation}
where 
\begin{equation}\label{lambda1}
    \lambda_{1}(\alpha; \sigma, x) = \mathcal{F}^{-1}\left[e^{-\sigma^{\alpha}|\kappa|^{\alpha}}\right] 
    =\frac{1}{\alpha|x|}
H_{2,2}^{1,1}\left[\frac{|x|}{\sigma}\left|\begin{array}{c
l}
    (1,1/\alpha), (1,1/2)\\
    (1,1), (1,1/2)
  \end{array}\right.\right]
\end{equation}
is the symmetric two-sided L\'{e}vy stable distribution which satisfies the scaling property, namely  (see, e.g., Ref.~\cite{ELukacs1970, KGorska2011})
$$\lambda_{1}(\alpha; \sigma, x) = \sigma^{-1} \lambda_{1}(\alpha; x \sigma^{-1}).$$

Here we used the Mellin-cosine transform of the Fox $H$-function~(\ref{cosine H}) and the relation between exponential and Fox $H$-function~(\ref{Hexp}). Another auxiliary function $\lambda_{2}(x)$ is defined through the inverse Fourier transform as follows
\begin{equation}\label{lambda2}
    \lambda_{2}(x)=\mathcal{F}^{-1}\left[e^{-\tilde{M}(\kappa)}\right],
\end{equation}
which form will depend on the kernel function $V(x)\equiv K(x)=M(x)\tau_{\gamma}^{-1}$.

\section{Presence of Stochastic Resetting}\label{Sec4}

In what follows, we will consider the process governed by eq.~(\ref{memory diffusion eq1}) under stochastic resetting. This means that the motion of a particle starting at $x = 0$ at time $t = 0$, described by the PDF $u(x,t)$, is renewed at random times following a distribution $r(t)$. Thus, the dynamics of the particle under resetting is described by the renewal equation for the PDF $u_{r}(x,t)$ in presence of resetting through the PDF $u(x,t)$ in absence of resetting, i.e.,~\cite{mendez_maso}
\begin{equation}\label{renewal_general}
    u_{r}(x,t)=R(t)u(x,t)+\int_{0}^{t}r(t')u(x,t')\,dt',
\end{equation}
where $R(t)=\int_{t}^{\infty}r(t')dt'$ is the probability that no renewal has taken place up to time $t$. Here we consider exponential (Poissonian) resetting for which $r(t)=re^{-rt}$, where $r$ is the resetting rate. Thus, $R(t)=e^{-rt}$, and the renewal equation has the form~\cite{maj_renew}
\begin{equation}\label{renewal}
    u_{r}(x,t)=e^{-rt}u(x,t)+\int_{0}^{t}re^{-rt'}u(x,t')\,dt'.
\end{equation}
Here we note that such generalised model under stochastic resetting but without nonlocal term was considered in~\cite{stwe}.

From the renewal equation, by Fourier-Laplace transform, it follows
\begin{equation}\label{renewal_general_fl}
    \tilde{\hat{U}}_{r}(\kappa,s)=\frac{s+r}{s}\tilde{\hat{U}}(\kappa,s+r)=\frac{s+r}{s}\frac{\hat{\gamma}(s+r)}{(s+r)\hat{\gamma}(s+r)+\mathcal{D}_{\alpha,\gamma}|\kappa|^{\alpha}+\tilde{K}(\kappa)}.
\end{equation}
Therefore, in the long time limit the system reaches the nonequilibrium stationary state (NESS)
\begin{equation}
    \tilde{U}_{r}^{st}(\kappa)=\lim_{t\rightarrow\infty}\tilde{U}_{r}(\kappa,t)=\lim_{s\rightarrow0}s\tilde{\hat{U}}_{r}(\kappa,s)=r\tilde{\hat{U}}(\kappa,r)=\frac{r\hat{\gamma}(r)}{r\hat{\gamma}(r)+\mathcal{D}_{\alpha,\gamma}|\kappa|^{\alpha}+\tilde{K}(\kappa)},
\end{equation}
which follows from the final value theorem $\lim_{t\rightarrow\infty}f(t)=\lim_{s\rightarrow0}s\hat{F}(s)$. Thus, we have
\begin{equation}\label{ness_general}
    u_{r}^{st}(x)=\mathcal{F}^{-1}\left[\frac{r\hat{\gamma}(r)}{r\hat{\gamma}(r)+\mathcal{D}_{\alpha,\gamma}|\kappa|^{\alpha}+\tilde{K}(\kappa)}\right]=\mathcal{F}^{-1}\left[\frac{1}{1+\frac{\mathcal{D}_{\alpha,\gamma}|\kappa|^{\alpha}+\tilde{K}(\kappa)}{r\hat{\gamma}(r)}}\right].
\end{equation}
In absence of nonlocal term the NESS becomes a Linnik distribution~\cite{stwe} since
\begin{equation}\label{ness_V0}
    u_{r}^{st}(x)=\mathcal{F}^{-1}\left[\frac{1}{1+\frac{\mathcal{D}_{\alpha,\gamma}}{r\hat{\gamma}(r)}|\kappa|^{\alpha}}\right],
\end{equation}
where
\begin{equation}
    Linnik(\alpha,x)=\mathcal{F}^{-1}\left[\frac{1}{1+A|\kappa|^{\alpha}}\right], \quad 0<\alpha<2.
\end{equation}
We note that for $\alpha=2$ it turns to the Laplace distribution, i.e.,
\begin{equation}
    u_{r}^{st}(x)=\mathcal{F}^{-1}\left[\frac{1}{1+\frac{\mathcal{D}_{2,\gamma}}{r\hat{\gamma}(r)}|\kappa|^{2}}\right]=\frac{1}{2}\sqrt{\frac{r\hat{\gamma}(r)}{\mathcal{D}_{2,\gamma}}}e^{-\sqrt{\frac{r\hat{\gamma}(r)}{\mathcal{D}_{2,\gamma}}}|x|},
\end{equation}
as it should be, see Ref.~\cite{stwe}.

\section{Long-range Nonlocal Term 
}\label{Sec5}

We consider the following long-range nonlocal term $V(x)=\frac{\vartheta}{2\cos\left(\pi\lambda/2\right)\Gamma(-\lambda)}|x|^{-\lambda-1}$, $0<\lambda<2$ and $\vartheta>0$, i.e., $\tilde{V}(\kappa)\equiv\tilde{K}(\kappa)=\vartheta|\kappa|^{\lambda}$ for initial condition $u(x,0)=\delta(x)$. Another form of the nonlocal term which satisfies the condition $\tilde{K}(\kappa)\rightarrow0$ for $\kappa\rightarrow0$ is of the form $V(x)=\frac{\vartheta}{2\cos\left(\pi\lambda/2\right)\Gamma(-\lambda)}|x-\bar{x}|^{-\lambda-1}$ since $\tilde{V}(\kappa)=\vartheta|\kappa|^{\lambda}e^{-\imath\kappa\bar{x}}$. For the PDF $u(x,t)$ in the Fourier-Laplace space in absence of resetting we obtain
\begin{equation}\label{memory diffusion eq laplace-fourier U |kappa|^lambda}
\tilde{\hat{U}}(\kappa,s)=\frac{\hat{\gamma}(s)}{s\hat{\gamma}(s)+\mathcal{D}_{\alpha,\gamma}|\kappa|^{\alpha}+\vartheta|\kappa|^{\lambda}}.
\end{equation}
Thus, the PDF reads
\begin{eqnarray}\label{memory diffusion eq space_time_U(x,t)}
u(x,t) &=&  \mathcal{F}^{-1}\left\{  \mathcal{L}^{-1}\left[  \frac{\hat{\gamma}(s)}{s \hat{\gamma}(s) + \mathcal{D}_{\alpha,\gamma}|\kappa|^{\alpha} +\vartheta|\kappa|^{\lambda}}\right] \right\} \nonumber\\
 & = &\mathcal{F}^{-1}\left[\int_0^\infty {\cal L}_\xi^{-1}\left[\frac{1}{s + \mathcal{D}_{\alpha,\gamma}|\kappa|^{\alpha} +\vartheta|\kappa|^{\lambda}}\right] {\cal L}_t^{-1}\left[\hat{\gamma}(s) e^{-\xi s \hat{\gamma}(s)}\right] d\xi\right] \nonumber\\
 & = &\int_0^\infty {\cal L}_\xi^{-1}\left[{\cal F}^{-1}\left[\frac{1}{s + \mathcal{D}_{\alpha,\gamma}|\kappa|^{\alpha} +\vartheta|\kappa|^{\lambda}}\right]\right]\, {\cal L}_t^{-1}\left[\hat{\gamma}(s) e^{-\xi s \hat{\gamma}(s)}\right] d\xi,
\end{eqnarray}
where we used Efros' theorem, see Appendix \ref{appB}. Therefore, we finally have
\begin{eqnarray}\label{10/11/24-1}
&&{\cal L}_\xi^{-1}\left[{\cal F}^{-1}\left[\frac{1}{s + \mathcal{D}_{\alpha,\gamma}|\kappa|^{\alpha} +\vartheta|\kappa|^{\lambda}}\right]\right] \nonumber\\
& &= \sum_{n=0}^\infty \frac{(-{\cal D}_{\alpha, \gamma})^n}{n!|x|^{1+\alpha n}}  {\cal L}_\xi^{-1}\left[s^{-1-n} 
H^{2, 1}_{2, 3}\left[\frac{|x|^\lambda}{\vartheta} s\left| 
\begin{array}{c}
(1,1), \big(\frac{2+\alpha n}{2}, \frac{\lambda}{2}\big) \\
(1 + \alpha n, \lambda), (1+n, 1), \big(\frac{2+\alpha n}{2}, \frac{\lambda}{2}\big)
\end{array}
\right. \right]
\right] \nonumber\\
&&= \sum_{n=0}^\infty \frac{(-{\cal D}_{\alpha, \gamma})^n}{n!} \frac{\xi^{n}}{|x|^{1+\alpha n}} 
H^{2, 1}_{3, 3}\left[\frac{|x|^\lambda}{\vartheta \xi}\left| 
\begin{array}{c}
(1,1), \big(\frac{2+\alpha n}{2}, \frac{\lambda}{2}\big), (n+1,1) \\
(1 + \alpha n, \lambda), (1+n, 1), \big(\frac{2+\alpha n}{2}, \frac{\lambda}{2}\big)
\end{array}
\right. \right]\nonumber\\
&&=\sum_{n=0}^\infty \frac{(-{\cal D}_{\alpha, \gamma})^n}{n!} |x|^{-1-\alpha n}\xi^{n} 
H^{1, 1}_{2, 2}\left[\frac{|x|^\lambda}{\vartheta \xi}\left| 
\begin{array}{c}
(1,1), \big(\frac{2+\alpha n}{2}, \frac{\lambda}{2}\big) \\
(1 + \alpha n, \lambda), \big(\frac{2+\alpha n}{2}, \frac{\lambda}{2}\big)
\end{array}
\right. \right],
\end{eqnarray}
where we used the inverse Laplace transform of the Fox $H$-function~(\ref{Laplace H}), and we applied the symmetry property~(\ref{prop_symm}) and reduction formula~(\ref{H_property02}) of the Fox $H$-function. This procedure by using the Efros' theorem actually corresponds to the subordination approach discussed in Remark~\ref{remark3}.

\begin{figure}[t!]
\centering{\includegraphics[width=6.5cm]{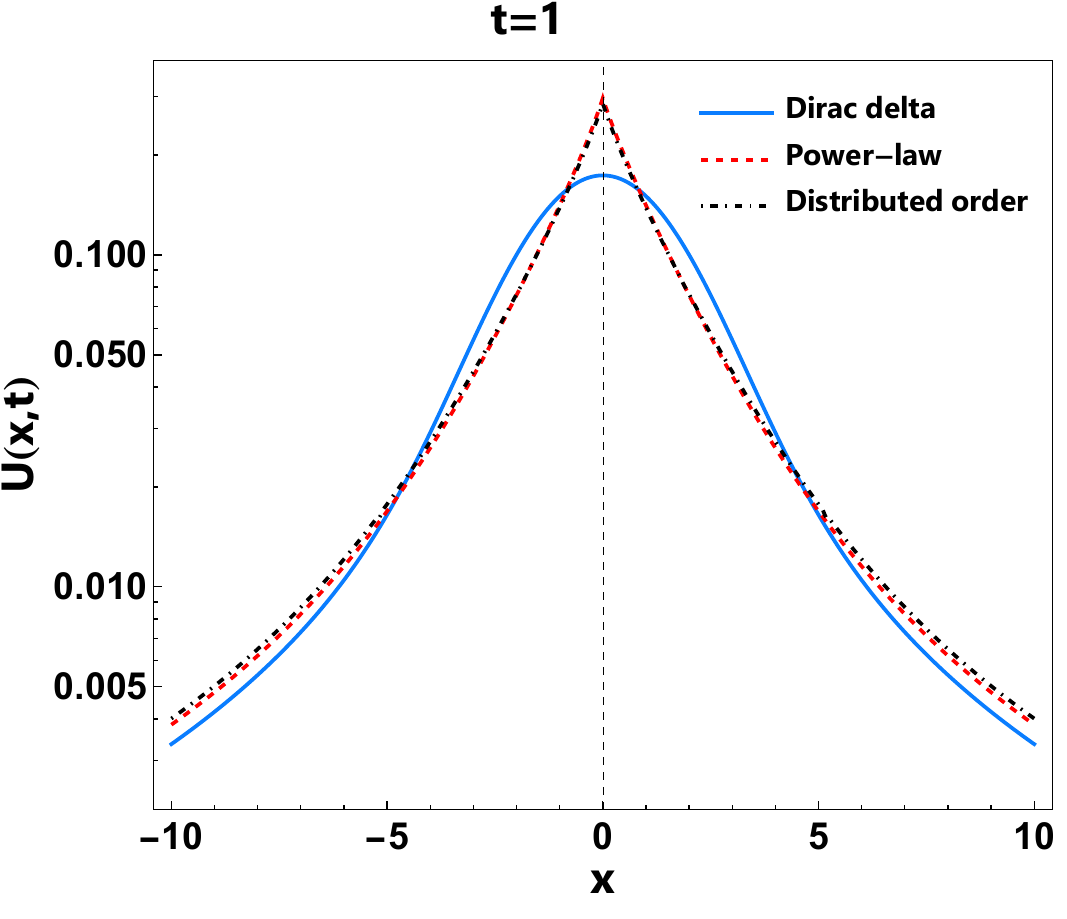} 
\includegraphics[width=6.5cm]{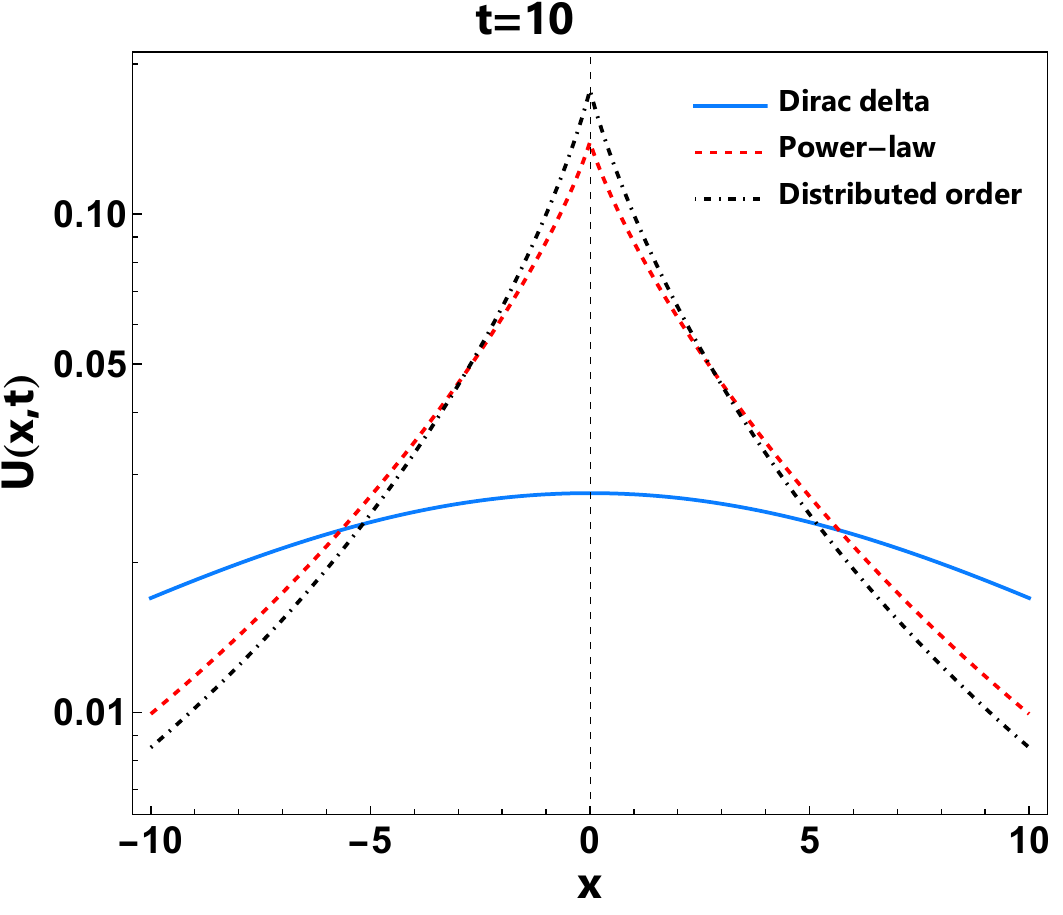}
\includegraphics[width=6.5cm]{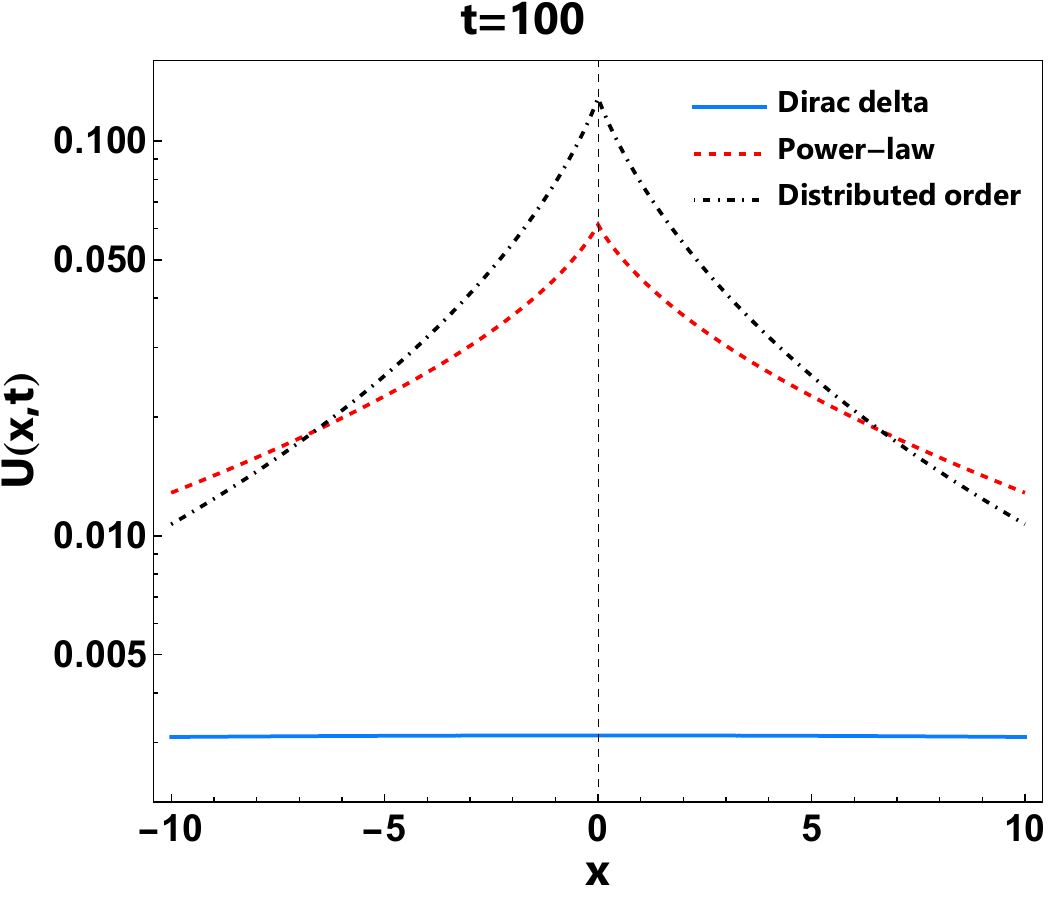}}
\caption{PDF~(\ref{memory diffusion eq space_time_U(x,t)}) at different times $t$ and three different types of kernels: $\gamma(t)=\delta (t)$ - blue solid line, $\gamma(t)=\frac{t^{-1/2}}{\Gamma(1/2)}$ - red dashed line, $\gamma(t)=\int_0^1 \frac{t^{-\mu}}{\Gamma(1-\mu)} d\mu$ - black dot-dashed line. We set $\mathcal{D}_{\alpha, \gamma}=1$, $\alpha=2$, $\theta=1$, $\lambda=1$.}\label{fig_PDF_no_reset}
\end{figure}

Graphical representation of the PDF~(\ref{memory diffusion eq laplace-fourier U |kappa|^lambda}) for different memory kernels is given in Fig.~\ref{fig_PDF_no_reset}. We plot the PDF by performing the numerical inverse Laplace transform together with the numerical inverse Fourier transform. This can be done, for example, by $\mathrm{NumericalInversion.m}$ package~\cite{mallet} from Wolfram Mathematica, where we combine the Stehfest method for numerical inverse Laplace transform with numerical inverse Fourier transform method, see also Appendix E in \cite{book_ws}. Here we note that we multiply expression~(\ref{renewal_ex_fl}) by a factor $\frac{1}{\sqrt{2\pi}}$ due to the symmetric definition of the Fourier and inverse Fourier transform in Mathematica.

\begin{remark}[{\bf Solution for Dirac delta memory kernel}]

For $\gamma(t)=\delta(t)$, one can find the solution from~(\ref{memory diffusion eq space_time_U(x,t)}), if we set $\hat{\gamma}(s)=1$. Moreover, from Remark~\ref{remark1} we find that the corresponding solution can be represented in the following form, as well
\begin{eqnarray}\label{rem1sol}
u(x,t)&=&\mathcal{F}^{-1}\left[e^{-\left(\mathcal{D}_{\alpha,\gamma}|\kappa|^{\alpha}+\vartheta|\kappa|^\lambda\right)t}\right]\nonumber\\& =& \int_{-\infty}^{\infty}\lambda_{1}(\alpha; (\mathcal{D}_{\alpha,1}t)^{1/\alpha}, x-\xi)\lambda_{1}(\lambda; (\vartheta t)^{1/\lambda}, \xi)\,d\xi\nonumber\\&=&\int_{-\infty}^{\infty}\frac{1}{\alpha|x-\xi|}
H_{2,2}^{1,1}\left[\frac{|x-\xi|}{(\mathcal{D}_{\alpha,1}t)^{1/\alpha}}\left|\begin{array}{c
l}
    (1,\frac{1}{\alpha}), (1,\frac{1}{2})\\
    (1,1), (1,\frac{1}{2})
  \end{array}\right.\right] \frac{1}{\lambda|\xi|}
H_{2,2}^{1,1}\left[\frac{|\xi|}{(\vartheta t)^{1/\lambda}}\left|\begin{array}{c
l}
    (1,\frac{1}{\lambda}), (1,\frac{1}{2})\\
    (1,1), (1,\frac{1}{2})
  \end{array}\right.\right]d\xi.
\end{eqnarray}

\end{remark}

\begin{figure}[t!]
\centering{\includegraphics[width=6.5cm]{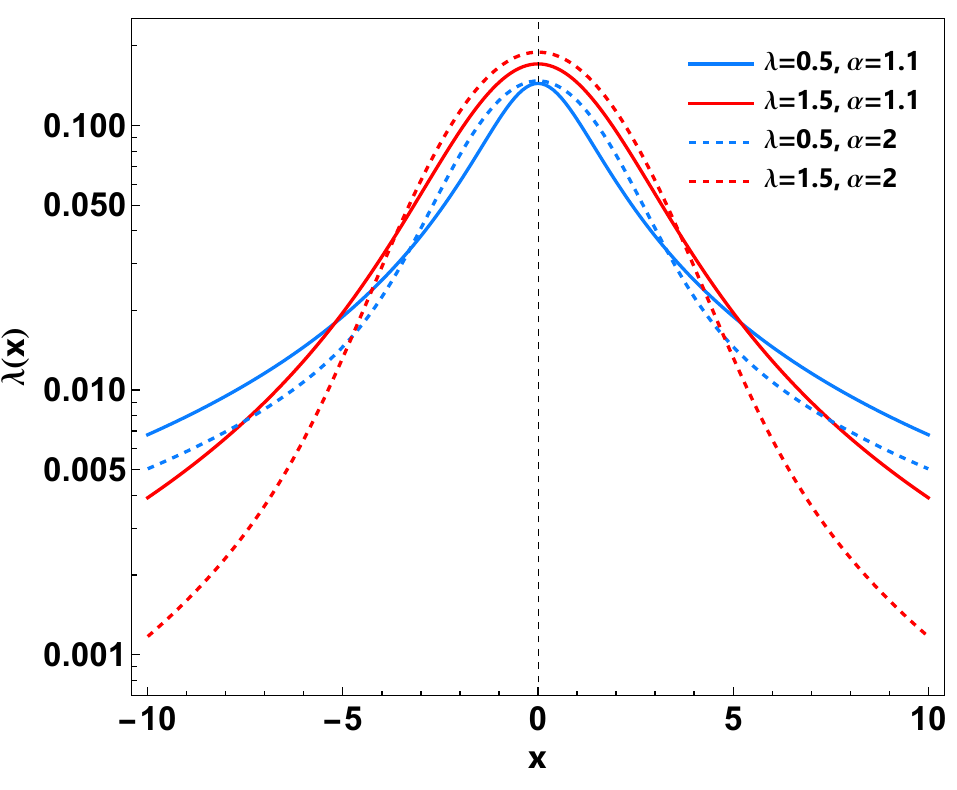} }
\caption{Jump length PDF~(\ref{lambda_composed_ex}) for different values of $\alpha$ and $\lambda$. We set $\bar{\vartheta}=1$ and $\mathcal{D}_{\alpha,\gamma}$=1.}\label{fig_jumpPDF}
\end{figure}

For this case one has decoupled CTRW with jump length PDF of form~(\ref{lambda_composed}), with $\lambda_1(\alpha; \sigma, x)$ and $\lambda_2(x) = \lambda_1(\lambda; \bar{\vartheta}^{1/\lambda}, x)$, $\bar{\vartheta}=\vartheta\tau_{\gamma}$, of form (\ref{lambda1}). Therefore,
\begin{eqnarray}\label{27/10/24-1}
    \lambda(x) = \int_{-\infty}^{\infty} \lambda_1(\alpha; \sigma, x - \xi)\, \lambda_1(\lambda; \bar{\vartheta}^{1/\lambda}, \xi)\,d\xi,
\end{eqnarray}
i.e.,
\begin{eqnarray}\label{lambda_composed_ex}
    \lambda(x) &= &\int_{-\infty}^{\infty}\frac{1}{\alpha|x-\xi|}
H_{2,2}^{1,1}\left[\frac{|x-\xi|}{\sigma}\left|\begin{array}{c
l}
    (1,\frac{1}{\alpha}), (1,\frac{1}{2})\\
    (1,1), (1,\frac{1}{2})
  \end{array}\right.\right]\frac{1}{\lambda|\xi|}
H_{2,2}^{1,1}\left[\frac{|\xi|}{\bar{\vartheta}^{1/\lambda}}\left|\begin{array}{c
l}
    (1,\frac{1}{\lambda}), (1,\frac{1}{2})\\
    (1,1), (1,\frac{1}{2})
  \end{array}\right.\right]d\xi.
\end{eqnarray}
This also can be represented by using generalized hypergeometric function~\cite{kg_jmp}. Graphical representation of the jump length PDF~(\ref{lambda_composed_ex}) for different values of $\alpha$ and $\lambda$ are given in Fig.~\ref{fig_jumpPDF}. 

Here we can find the PDF in presence of resetting by using the renewal equation in Fourier-Laplace space~(\ref{renewal_general_fl}), which reads
\begin{eqnarray}\label{renewal_ex_fl}
    \tilde{\hat{U}}_{r}(\kappa,s)=\frac{s+r}{s}\frac{\hat{\gamma}(s+r)}{(s+r)\hat{\gamma}(s+r)+\mathcal{D}_{\alpha,\gamma}|\kappa|^{\alpha}+\vartheta|\kappa|^{\lambda}}.
\end{eqnarray}

\begin{figure}[t!]   \centering{\includegraphics[width=6.5cm]{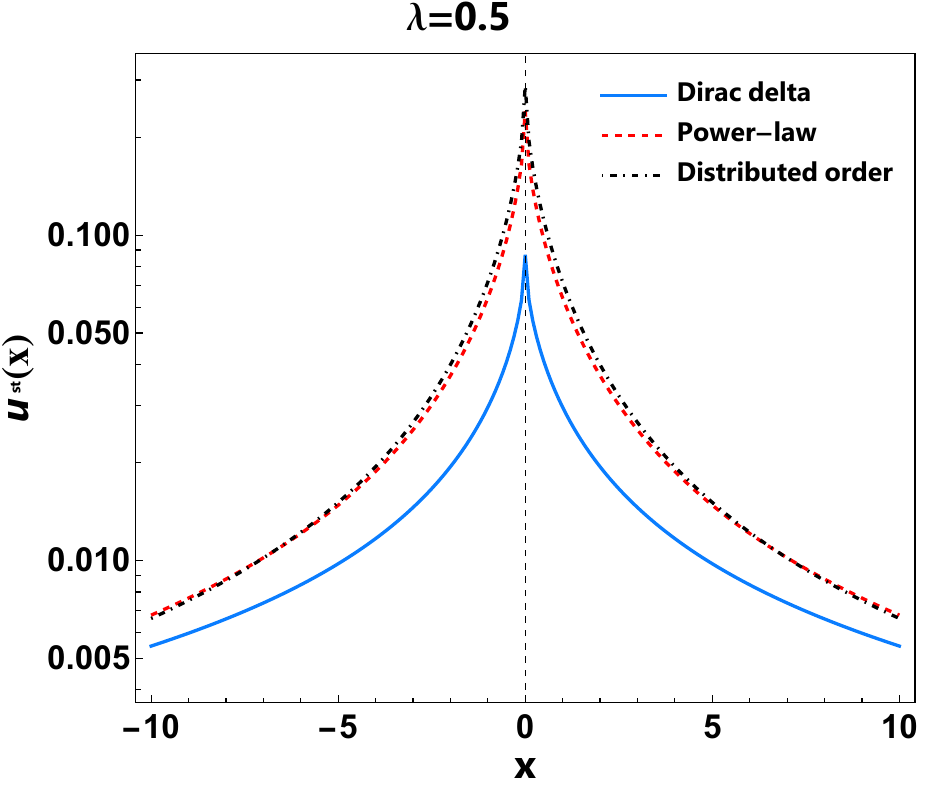} 
 \includegraphics[width=6.5cm]{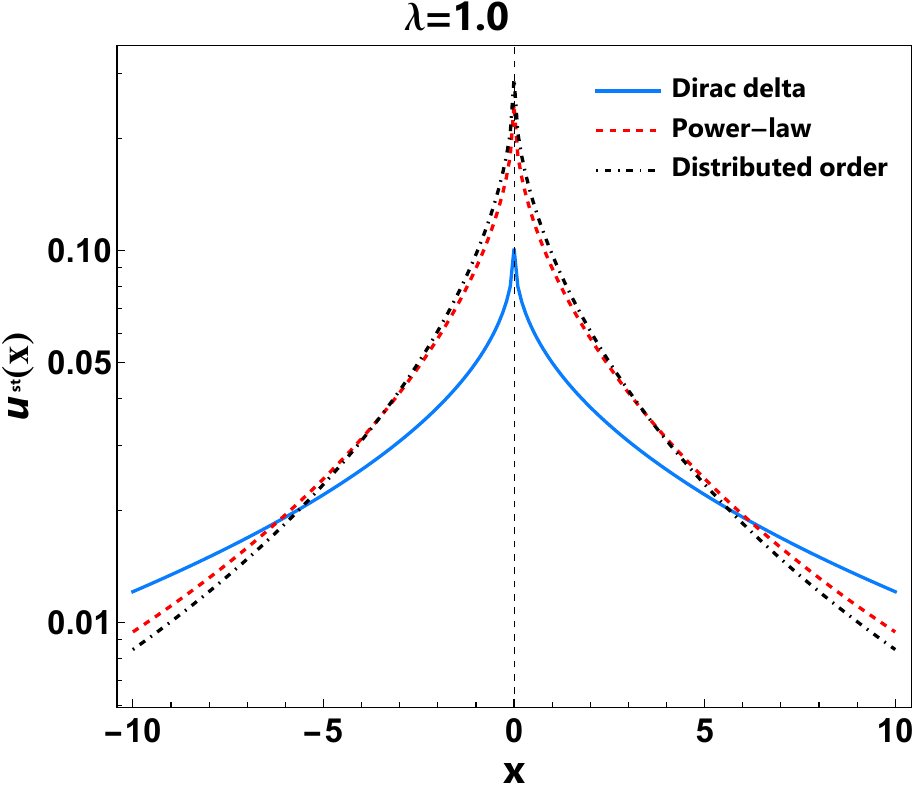}
 \includegraphics[width=6.5cm]{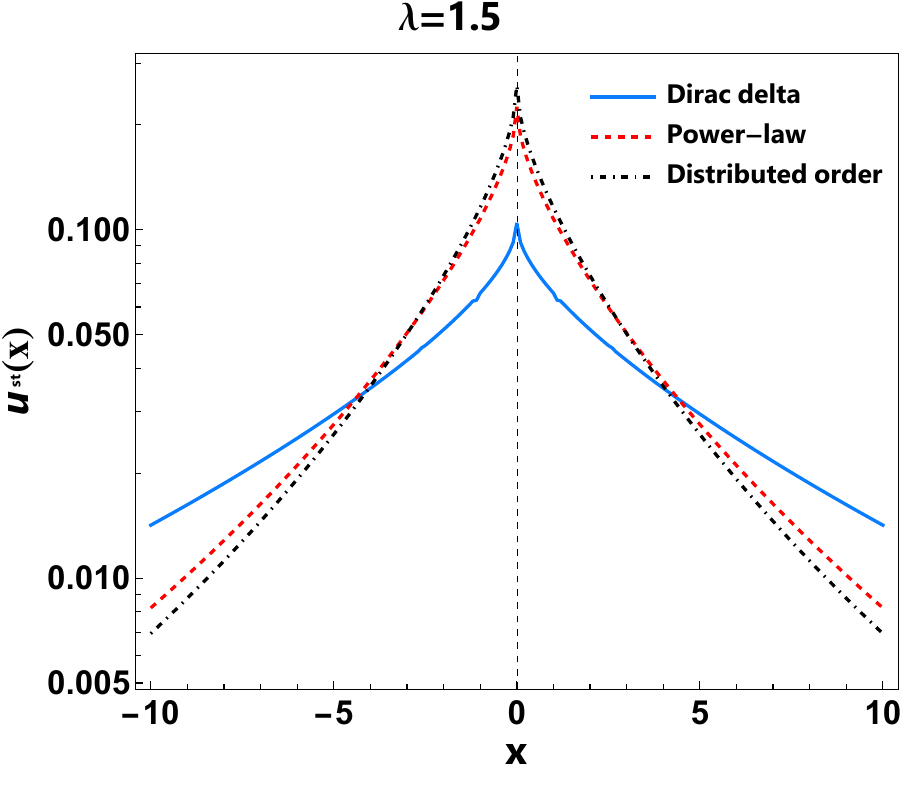}}
\caption{NESS~(\ref{ness_general_nolocal}) for various $\lambda$ and three different types of kernel: NESS~(\ref{ness_general_nolocal_delta}) - blue solid line, NESS~(\ref{ness_general_nolocal_fract}) - red dashed line, NESS~(\ref{ness_general_nolocal_distr}) - black dot-dashed line. We set $\vartheta=1$, $\mathcal{D}_{\alpha,\gamma}=1$, $\mu=0.5$, $r=0.1$, $\alpha=1.5$.}\label{fig_NESS_gamma}
\end{figure}

From eq.~(\ref{renewal_ex_fl}), we find the NESS in terms of infinite series in Fox $H$-function, which is reached in the long time limit, 
\begin{eqnarray}\label{ness_general_nolocal}
    u^{st}(x)&=&\mathcal{F}^{-1}\left[\frac{1}{1+\frac{\mathcal{D}_{\alpha,\gamma}|\kappa|^{\alpha}+\vartheta|\kappa|^{\lambda}}{r\hat{\gamma}(r)}}\right]=\frac{1}{\pi}\int_{0}^{\infty}\frac{\cos(\kappa x)}{1+\frac{\mathcal{D}_{\alpha,\gamma}\kappa^{\alpha}+\vartheta\kappa^{\lambda}}{r\hat{\gamma}(r)}}d\kappa \nonumber\\
    & =& \sum_{n=0}^\infty \frac{(-{\cal D}_{\alpha, \gamma})^n}{\big(r \hat{\gamma}(r)\big)^n\, n!\, |x|^{\alpha n +1}} H^{2, 1}_{2, 3}\left[\frac{r \hat{\gamma}(r)|x|^\lambda}{\vartheta} \left|
    \begin{array}{c}
    (1, 1), \big(\frac{2+\alpha n}{2}, \frac{\lambda}
{2}\big) \\    
(\alpha n +1, \lambda), (1+n, 1), \big(\frac{2+\alpha n}{2}, \frac{\lambda}
{2}\big)
    \end{array}
    \right. \right].
\end{eqnarray}
This NESS for $\gamma(t)=\delta(t)$ becomes
\begin{equation}\label{ness_general_nolocal_delta}
    u^{st}(x)
    = \sum_{n=0}^\infty \frac{(-{\cal D}_{\alpha, \gamma})^n}{r^n\, n!\, |x|^{\alpha n +1}} H^{2, 1}_{2, 3}\left[\frac{r|x|^\lambda}{\vartheta} \left|
    \begin{array}{c}
    (1, 1), \big(\frac{2+\alpha n}{2}, \frac{\lambda}
{2}\big) \\    
(\alpha n +1, \lambda), (1+n, 1), \big(\frac{2+\alpha n}{2}, \frac{\lambda}
{2}\big)
    \end{array}
    \right. \right].
\end{equation}
For $\gamma(t)=\frac{t^{-\mu}}{\Gamma(1-\mu)}$, $0<\mu<1$, it reads
\begin{equation}\label{ness_general_nolocal_fract}
    u^{st}(x)
    = \sum_{n=0}^\infty \frac{(-{\cal D}_{\alpha, \gamma})^n}{r^{\mu n}\, n!\, |x|^{\alpha n +1}} H^{2, 1}_{2, 3}\left[\frac{r^{\mu}|x|^\lambda}{\vartheta} \left|
    \begin{array}{c}
    (1, 1), \big(\frac{2+\alpha n}{2}, \frac{\lambda}
{2}\big) \\    
(\alpha n +1, \lambda), (1+n, 1), \big(\frac{2+\alpha n}{2}, \frac{\lambda}
{2}\big)
    \end{array}
    \right. \right]
\end{equation}
while for uniformly distributed order memory kernel it has the form
\begin{equation}\label{ness_general_nolocal_distr}
    u^{st}(x)
    = \sum_{n=0}^\infty \frac{(-{\cal D}_{\alpha, \gamma})^n}{\left[\frac{(r-1)}{\log{r}}\right]^{n} n!\, |x|^{\alpha n +1}} H^{2, 1}_{2, 3}\left[\frac{(r-1)|x|^\lambda}{\vartheta\log{r}} \left|
    \begin{array}{c}
    (1, 1), \big(\frac{2+\alpha n}{2}, \frac{\lambda}
{2}\big) \\    
(\alpha n +1, \lambda), (1+n, 1), \big(\frac{2+\alpha n}{2}, \frac{\lambda}
{2}\big)
    \end{array}
    \right. \right].
\end{equation}
Graphical representation of the NESS~(\ref{ness_general_nolocal}) for different memory kernels and various $\lambda$ is given in Fig.~\ref{fig_NESS_gamma}.

\begin{remark}[\textbf{Special case}]
We note that some special cases can be solved exactly. For example, for the case with $\alpha=2$ and $\lambda=1$, one finds 
\begin{eqnarray}
    u_{r}^{st}(x)&=&\mathcal{F}^{-1}\left[\frac{1}{1+\frac{\mathcal{D}_{2,\gamma}\kappa^{2}+\vartheta|\kappa|}{r\hat{\gamma}(r)}}\right]=\frac{2}{2\pi}\int_{0}^{\infty}\frac{\cos(\kappa x)}{1+\frac{\vartheta}{r\hat{\gamma}(r)}\kappa+\frac{\mathcal{D}_{2,\gamma}}{r\hat{\gamma}(r)}\kappa^{2}}d\kappa\nonumber\\&=&\frac{\mathrm{ci}(k_2x)\cos(k_2x)-\mathrm{ci}(k_1x)\cos(k_1x)+\mathrm{si}(k_2x)\sin(k_2x)-\mathrm{si}(k_1x)\sin(k_1x)}{\pi\frac{\mathcal{D}_{2,\gamma}}{r\hat{\gamma}(r)}(k_2-k_1)},\nonumber\\
\end{eqnarray}
where $$k_{1,2}=\frac{\vartheta\mp\sqrt{\vartheta^2-\mathcal{D}_{2,\gamma}r\hat{\gamma}(r)}}{2\mathcal{D}_{2,\gamma}},$$ while $\mathrm{si}(z)=-\int_{z}^{\infty}\frac{sin{y}}{y}dy$ and $\mathrm{ci}(z)=-\int_{z}^{\infty}\frac{\cos{y}}{y}dy$ are integral sine and integral cosine, see Refs.~\cite{prudnikov,palyulin}. Here we note that in Wolfram Mathematica it is used the following function $\mathrm{Si}(z)=\int_{0}^{z}\frac{\sin{y}}{y}dy=\frac{\pi}{2}+\mathrm{si}(z)$, while $\mathrm{Ci}(z)$ corresponds to $\mathrm{ci}(z)$.
\end{remark}

\section{Summary and Future Directions}\label{sec_summary}

In this work we analysed generalised diffusion equation with memory kernel and Riesz space fractional derivative in presence of a nonlocal term. We give a general solution of the problem and derive the same equation within the CTRW theory. We showed that the solution of the considered equation can be found if one knows the solution of the corresponding equation without memory kernel by using the subordination approach. We also consider the process with stochastic resetting and showed that the system reaches a nonequilibrium stationary state in the long time limit. We applied the obtained result to the case with long range nonlocal term and different forms of the memory kernel (Dirac delta, power-law and distributed order memory kernel). We showed that for such long range nonlocal term the jump length PDF in the CTRW model can be represented as a convolution integral of two symmetric two-sided L\'evy stable distributions, while the nonequilibrium stationary state reached in presence of resetting is a generalisation of the Linnik and Laplace distributions. 

An interesting generalisation of the considered model~(\ref{memory diffusion eq1}) is the following equation
\begin{equation}\label{memory diffusion eq}
\int_{0}^{t}\gamma(t-\tau)\frac{\partial}{\partial\tau}u(x,\tau)\,d\tau=\mathcal{D}_{\alpha,\gamma}\frac{\partial^{\alpha}}{\partial |x|^{\alpha}}u(x,t)-\int_{0}^{t}\left[\int_{-\infty}^{\infty}V(x-\xi,t-\tau)u(\xi,\tau)\,d\xi\right]d\tau,\nonumber\\
\end{equation}
where the nonlocal term is of form $V(x,t)=K(x)\varphi(t)$, where the kernel function $V(x)\equiv K(x)$ is such that $\tilde{K}(\kappa=0)=0$. Therefore, from Eq.~(\ref{memory diffusion eq}) in the Fourier-Laplace space we have
\begin{equation}\label{memory diffusion eq laplace-fourier U varphi(t)}
\tilde{\hat{U}}(\kappa,s)=\frac{\hat{\gamma}(s)}{s\hat{\gamma}(s)+\mathcal{D}_{\alpha,\gamma}|\kappa|^{\alpha}+\tilde{K}(\kappa)\hat{\varphi}(s)}\tilde{U}(\kappa,0).
\end{equation}
Finding the corresponding CTRW model for this generalised equation with nonlocal term we leave for future research since due to the term $\tilde{K}(\kappa)\hat{\varphi}(s)$ in eq.~(\ref{memory diffusion eq laplace-fourier U varphi(t)}) one can not use the uncoupled CTRW model discussed in this work. Instead, it should be treated by the coupled CTRW model, which means that the jumps are penalised by waiting times. 

Future investigation of the generalised diffusion processes with nonlocal term will be related to the corresponding random search process~\cite{jstat,palyulin,pre}, as well as processes with time-dependent resetting~\cite{97}, non-instantaneous~\cite{98,98_1} and space–time coupled returns~\cite{99}, processes in presence of resetting in an interval~\cite{100_1,101} and bounded in complex potential~\cite{102}, as well as discrete space–time resetting models~\cite{103}, resetting to multiple~\cite{multiple} and random positions~\cite{random}. The corresponding processes of quantum motion with long range nonlocal terms is also a topics worth of investigation.

\begin{acknowledgments}
{PT, IP, LK, and TS acknowledge financial support
by the German Science Foundation (DFG, Grant number ME 1535/12-1) and by the bilateral Macedonian-Austrian project No. 20-667/10 (WTZ MK03/2024). PT, LK, and TS are also supported by the Alliance of International Science Organizations (Project No. ANSO-CR-PP-2022-05). TS was supported by the Alexander von Humboldt Foundation. KG acknowledges the financial support provided under the NCN Research Grant Preludium Bis 2 No. UMO-2020/39/O/ST2/01563.}
\end{acknowledgments}

\appendix

\section{Fox $H$-function}

The Fox $H$-function (or simply $H$-function) is defined by the
following Mellin-Barnes integral~\cite{saxena book}
\begin{eqnarray}\label{H_integral}
H_{p,q}^{m,n}\left[z\left|\begin{array}{c l}
    (a_1,A_1),...,(a_p,A_p)\\
    (b_1,B_1),...,(b_q,B_q)
  \end{array}\right.\right]=H_{p,q}^{m,n}\left[z\left|\begin{array}{c l}
    (a_p,A_p)\\
    (b_q,B_q)
  \end{array}\right.\right]=\frac{1}{2\pi\imath}\int_{\Omega}\theta(s)z^{s}\,ds,
\end{eqnarray}
where
$$\theta(s)=\frac{\prod_{j=1}^{m}\Gamma(b_j-B_js)\prod_{j=1}^{n}\Gamma(1-a_j+A_js)}{\prod_{j=m+1}^{q}\Gamma(1-b_j+B_js)\prod_{j=n+1}^{p}\Gamma(a_j-A_js)},$$
$0\leq n\leq p$, $1\leq m\leq q$, $a_i,b_j \in C$, $A_i,B_j \in
R^{+}$, $i=1,...,p$, $j=1,...,q$. The contour $\Omega$ starting at
$c-\imath\infty$ and ending at $c+\imath\infty$ separates the poles
of the function $\Gamma(b_j+B_js)$, $j=1,...,m$ from those of the
function $\Gamma(1-a_i-A_is)$, $i=1,...,n$. The expansion for the $H$-function~(\ref{H_integral}) is given by~\cite{saxena book}
\begin{eqnarray}\label{H_expansion}
H_{p,q}^{m,n}\left[z\left|\begin{array}{c l}
    (a_p,A_p)\\
    (b_q,B_q)
  \end{array}\right.\right]=\sum_{h=1}^{m}\sum_{k=0}^{\infty}&\frac{\prod_{j=1, j\neq
h}^{m}\Gamma\left(b_j-B_j\frac{b_h+k}{B_h}\right)\prod_{j=1}^{n}\Gamma\left(1-a_j+A_j\frac{b_h+k}{B_h}\right)}{\prod_{j=m+1}^{q}\Gamma\left(1-b_j+B_j\frac{b_h+k}{B_h}\right)\prod_{j=n+1}^{p}\Gamma\left(a_j-A_j\frac{b_h+k}{B_h}\right)}\frac{(-1)^kz^{(b_h+k)/B_h}}{k!B_h}.
\end{eqnarray}

The one parameter Mittag-Leffler function (\ref{one parameter ML}) is a special case of
Fox $H$-function~\cite{saxena book}
\begin{eqnarray}\label{HML}
E_{\alpha}(-z)=H_{1,2}^{1,1}\left[z\left|\begin{array}{cc}
    (0,1)\\
    (0,1),(0,\alpha)
  \end{array}\right.\right],
\end{eqnarray}
as well as the exponential function,
\begin{eqnarray}\label{Hexp}
e^{-z} =H_{0,1}^{1,0}\left[z\left|\begin{array}{cc}
    - \\
    (0,1)
  \end{array}\right.\right].
\end{eqnarray}

The $H$-function has the following properties~\cite{saxena book}:

\begin{proposition}[{\bf Symmetry property}]\label{prop_symm}

The Fox $H$-function is symmetric in the following pairs $(a_1,A_1),\dots,(a_n,A_n)$, as well as in the pairs $(a_{n+1},A_{n+1}),\dots,(a_p,A_p)$. The Fox $H$-function is symmetric also in the pairs $(b_1,B_1),\dots,(b_m,B_m)$, as well as in the pairs $(b_{m+1},B_{m+1}),\dots,(b_q,B_q)$.
\end{proposition}

\begin{proposition}[{\bf Reduction formulas}]\label{H_reduction_prop}
The following reduction formulas
\begin{eqnarray}\label{H_property0}
&&H_{p,q}^{m,n}\left[z\left|\begin{array}{c l}
    (a_1,A_1),...,(a_p,A_p)\\
    (b_1,B_1),...,(b_{q-1},B_{q-1}),(a_1,A_1)
  \end{array}\right.\right]=H_{p-1,q-1}^{m,n-1}\left[z\left|\begin{array}{c l}
    (a_2,A_2),...,(a_p,A_p)\\
    (b_1,B_1),...,(b_{q-1},B_{q-1})
  \end{array}\right.\right],
\end{eqnarray}
and
\begin{eqnarray}\label{H_property02}
&&H_{p,q}^{m,n}\left[z\left|\begin{array}{c l}
    (a_1,A_1),...,(a_{p-1},A_{p-1}),(b_1,B_1)\\
    (b_1,B_1),...,(b_q,B_q)
  \end{array}\right.\right]=H_{p-1,q-1}^{m-1,n}\left[z\left|\begin{array}{c l}
    (a_1,A_1),...,(a_{p-1},A_{p-1})\\
    (b_2,B_2),...,(b_{q},B_{q})
  \end{array}\right.\right],
\end{eqnarray}
are valid for $n\geq1$, $q>m$.
\end{proposition}

\begin{proposition}
For $\delta>0$, it is valid 

\begin{eqnarray}\label{H_property}
H_{p,q}^{m,n}\left[z^{\delta}\left|\begin{array}{c l}
    (a_p,A_p)\\
    (b_q,B_q)
  \end{array}\right.\right]=\frac{1}{\delta}\, H_{p,q}^{m,n}\left[z\left|\begin{array}{c l}
    (a_p,A_p/\delta)\\
    (b_q,B_q/\delta)
  \end{array}\right.\right].
\end{eqnarray}

\end{proposition}

\begin{proposition} The following property also holds true

\begin{eqnarray}\label{H_property2}
z^{\sigma}H_{p,q}^{m,n}\left[z\left|\begin{array}{c l}
    (a_p,A_p)\\
    (b_q,B_q)
  \end{array}\right.\right]=H_{p,q}^{m,n}\left[z\left|\begin{array}{c l}
    (a_p+\sigma A_p,A_p)\\
    (b_q+\sigma B_q,B_q)
  \end{array}\right.\right].
\end{eqnarray}
\end{proposition}

The Mellin-cosine transform of the $H$-function is given by~\cite{saxena book}

\begin{eqnarray}\label{cosine H}
&&\int_{0}^{\infty}k^{\rho-1}\cos(kx)H_{p,q}^{m,n}\left[ak^{\delta}\left|\begin{array}{c
l}
    (a_p,A_p)\\
    (b_q,B_q)
  \end{array}\right.\right]dk=\frac{\pi}{x^\rho}H_{q+1,p+2}^{n+1,m}\left[\frac{x^\delta}{a}\left|\begin{array}{c
l}
     (1-b_q,B_q), (\frac{1+\rho}{2}, \frac{\delta}{2})\\
    (\rho,\delta), (1-a_p,A_p), (\frac{1+\rho}{2},\frac{\delta}{2})
  \end{array}\right.\right],
\end{eqnarray}
where $\Re\left(\rho+\delta \min_{1\leq j\leq
m}\left(\frac{b_j}{B_j}\right)\right)>1$, $x^\delta>0$,
$\Re\left(\rho+\delta \max_{1\leq j\leq
n}\left(\frac{a_j-1}{A_j}\right)\right)<\frac{3}{2}$,
$|\arg(a)|<\pi\theta/2$, $\theta>0$,
$\theta=\sum_{j=1}^{n}A_j-\sum_{j=n+1}^{p}A_j+\sum_{j=1}^{m}B_j-\sum_{j=m+1}^{q}B_j$.

The inverse Laplace transform of the Fox $H$-function reads~\cite{saxena book}
\begin{eqnarray}\label{Laplace H}
\mathcal{L}^{-1}\left[s^{-\rho}H_{p,q}^{m,n}\left[as^{\sigma}\left|\begin{array}{c
l}
    (a_p,A_p)\\
    (b_q,B_q)
  \end{array}\right.\right]\right]
  =t^{\rho-1}H_{p+1,q}^{m,n}\left[\frac{a}{t^{\sigma}}\left|\begin{array}{c
l}
    (a_p,A_p), (\rho,\sigma)\\
    (b_q,B_q)
  \end{array}\right.\right].
\end{eqnarray}

\section{Efros' Theorem}\label{appB}

Efros' theorem \cite{AMEfros35, LWlodarski52, UGraf04, KGorska12a} generalizes the convolution theorem for the Laplace transform. It states as follows:

\begin{theorem}\label{t1}
If $\hat{G}(s)$ and $\hat{q}(s)$ are analytic functions, and $\mathcal{L}[h(x, \xi)] = \hat{h}(x, s)$ as well as $\mathcal{L}[f(\xi, t)] = \int_{0}^{\infty} f(\xi, t) e^{-z t} d t = \hat{G}(s) e^{-\xi \hat{q}(s)}$ exist, then 
\begin{equation}
\hat{G}(s)\hat{h}(x,\hat{q}(s)) = \int_{0}^{\infty}\left[\int_{0}^{\infty} f(\xi, t) h(x, \xi) d\xi\right] e^{- s t} d t.
\end{equation}
\end{theorem}


\end{document}